\documentclass[twocolumn,showpacs,preprintnumbers,amsmath,amssymb,superscriptaddress,prl,longbibliography]{revtex4-1}

\usepackage{graphicx}
\usepackage{xcolor}

\allowdisplaybreaks

\begin{document}

\title{Anomalously long lifetimes in two-dimensional Fermi liquids}

\author{Johannes Hofmann}
\email{johannes.hofmann@physics.gu.se}
\affiliation{Department of Physics, Gothenburg University, 41296 Gothenburg, Sweden}

\author{Ulf Gran}
\email{ulf.gran@chalmers.se}
\affiliation{Department of Physics, Chalmers University of Technology, 41296 Gothenburg, Sweden}

\date{\today}

\begin{abstract}
A precise characterization of the recently discovered crossover to hydrodynamic transport in electron liquids, and in particular of a conjectured exotic  odd-parity transport regime, requires a full solution of the Fermi-liquid collision integral at all temperatures beyond state-of-the-art analytic leading-logarithmic approximations. We develop a general basis expansion of the linearized collision integral applicable at all temperatures, and employ this to determine the decay rates of Fermi surface perturbations in two-dimensional electron liquids. In particular, we provide exhaustive data on the decay rates for a new odd-parity regime, in which parity-even Fermi surface deformations decay rapidly but odd-parity modes are anomalously long lived. We find that this transport regime exists at fairly large temperatures \mbox{$T \lesssim 0.1 \, T_F$}, putting this regime within the reach of current experiments.
\end{abstract}

\maketitle

Signatures of hydrodynamic transport have recently been reported in a number of two-dimensional materials such as graphene~\cite{bandurin16,crossno16,
krishnakumar17,berdyugin19}, bilayer graphene~\cite{nam17,bandurin18}, GaAs~\cite{dejong95,buhmann02,
braem18}, WP$_2$~\cite{gooth18}, and PdCoO$_2$~\cite{moll16}. 
Samples used are very clean, such that at intermediate temperatures electron-electron scattering---the rate of which goes as \mbox{$\gamma \sim (T/T_F)^2$} at small temperatures  with logarithmic corrections~\cite{chaplik71,hodges71,giuliani82,zheng96,dassarma21}, where $T_F$ is the Fermi temperature---sets the dominant relaxation mechanism over impurity and phonon scattering (which dominate at low and  high temperatures, respectively)~\cite{lucas18}. In this case, excitations not associated with conserved quantities (charge and current) relax very quickly due to electron interactions, giving rise to a collective hydrodynamic description. Indeed, hydrodynamic transport is generally expected in many two-dimensional materials~\cite{bandurin16,bandurin18,ahn22}. On a theoretical level, the crossover to a hydrodynamic regime is captured by a kinetic quasiparticle description~\cite{pines66,smith89,baym04,giuliani05,Lucas:2018kwo}.

However, this standard argument for the rapid decay of nonconserved modes at finite temperature is incomplete: It only applies to a subset of all excitations (dubbed ``even-parity'' modes), and there exist long-lived modes (dubbed ``odd-parity'' modes) that decay much more slowly with a predicted rate $\gamma' \sim (T/T_F)^4$ at low temperatures~\cite{ledwith17,ledwith19}. This implies that 2D materials should posses a new exotic transport regime distinct form the hydrodynamic one in which even-parity modes decay quickly, but odd-parity modes are still nearly collisionless. This new regime has so far been overlooked in experiments as most observables (such as the resistivity)  couple to conserved modes or involve a superposition of odd- and even-parity modes and will be sensitive to the strongest decay channel. Nevertheless, signatures of this odd-even effect in the decay rates are predicted as a shear-thinning viscosity~\cite{ledwith19b}, which leads to a modification of the hydrodynamic Poiseuille flow profile in a narrow channel that might in fact have been observed already~\cite{sulpizio19}, or the emergence of new transverse diffusive collective modes, which do not couple to the charge~\cite{hofmann22}.
The description of such a transport regime---in terms of transport coefficients, collective modes, or decay rates---requires a full numerical solution at all temperatures of the Fermi liquid equations that govern the electron liquid. To develop such a general formalism, and use it to obtain a complete characterization of the decay rates for all modes and all temperatures, is the purpose of this Letter.

\begin{figure}[b!]\vspace{-0.5cm}
\scalebox{1}{\includegraphics{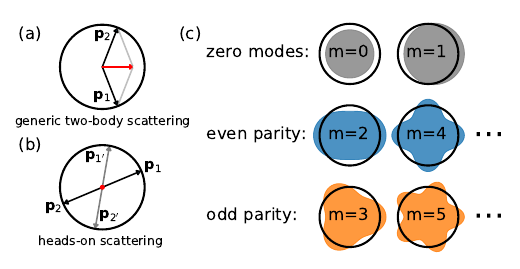}}
\caption{(a) Generic two-body scattering process. (b) Head-on scattering that relaxes even-parity deformations. (c) Fermi surface deformation of zero modes (gray), and the first even-parity  (blue) and odd-parity modes (orange). $m$ is the angular mode number.}
\label{fig:1}
\end{figure}

To understand the emergent disparity of lifetimes on a qualitative level, consider the relaxation of Fermi surface deformations in a kinetic picture at low temperatures [we use a parabolic dispersion $\varepsilon({\bf p}) = p^2/2m^*$ ($m^*$ is an effective band mass) with circular Fermi surface]. A typical electron-electron scattering process \mbox{${\bf p}_1 + {\bf p}_2 \to {\bf p}_{1'} + {\bf p}_{2'}$} [Fig.~\ref{fig:1}(a)] involves momenta close to the Fermi momentum (black circle), where black arrows indicate ${\bf p}_1$ and ${\bf p}_2$ and the red arrow is the total momentum. Given momentum conservation, only forward (\mbox{${\bf p}_{1'}={\bf p}_{1},{\bf p}_{2'}={\bf p}_{2}$}) or exchange scattering (\mbox{${\bf p}_{1'}={\bf p}_{2},{\bf p}_{2'}={\bf p}_{1}$}) is possible, neither of which relax the (spin unresolved) quasiparticle distribution. The only exception is head-on scattering with zero center-of-mass momentum [Fig.~\ref{fig:1}(b)], which in a microscopic calculation gives rise to the \mbox{$\gamma \sim (T/T_F)^2$} scaling (with a logarithmic enhancement for nearly collinear scattering~\cite{laikhtman92}). However, such processes relax opposite points on the Fermi surface in equal manner. This then implies a parity-dependence of the lifetime [Fig.~\ref{fig:1}(c)]: 
In a Fourier mode expansion of Fermi surface deformations, in addition to the conserved (zero) modes (gray), even-parity modes (blue) decay quickly at finite temperature in the manner described above while odd-parity modes (orange) decay slowly by other means. Of course, a full calculation cannot assume a rigid Fermi surface deformation but must include the softening of the Fermi-Dirac distribution at finite temperature.

In this Letter, we provide a general framework to solve the linearized collision integral for electron-electron scattering in two-dimensional Fermi liquids at all temperatures, which is necessary to determine the linear response and the lifetime of excitations. 
The full collision integral $\hat{\cal J}$ describes the rate of change of the quasiparticle distribution function $f(t,{\bf p})$ due to
 momentum-changing two-body collisions with other particles:
\begin{widetext}
\begin{eqnarray}
&&
\hat{\cal J}[f(t, {\bf p}_1)]
= - \frac{2\pi}{\hbar} \, \int \frac{d{\bf p}_2}{(2\pi\hbar)^2} \int \frac{d{\bf p}_{1'}}{(2\pi\hbar)^2} \int \frac{d {\bf p}_{2'}}{(2\pi\hbar)^2} \, \Bigl|\langle {\bf p}_1, {\bf p}_2 | V | {\bf p}_{1'}, {\bf p}_{2'} \rangle \Bigr|^2 \delta\bigl(\varepsilon_{{\bf p}_1} + \varepsilon_{{\bf p}_2} - \varepsilon_{{\bf p}_{1'}} - \varepsilon_{{\bf p}_{2'}}\bigr) \nonumber \\
&& \times (2\pi\hbar)^2 \delta\bigl({\bf p}_1 + {\bf p}_2 - {\bf p}_{1'} - {\bf p}_{2'}\bigr) \Bigl[ f({\bf p}_1) f({\bf p}_2) \bigl(1 - f({\bf p}_{1'})\bigr) \bigl(1 - f({\bf p}_{2'})\bigr) 
- 
f({\bf p}_{1'}) f({\bf p}_{2'}) \bigl(1 - f({\bf p}_{1})\bigr) \bigl(1 - f({\bf p}_{2})\bigr)\Bigr] . \label{eq:collision}
\end{eqnarray}
\end{widetext}
The first term in the square brackets describes a loss of quasiparticles with momentum ${\bf p}_1$ due to a scattering process ${\bf p}_1 + {\bf p}_2 \to {\bf p}_{1'} + {\bf p}_{2'}$ (with matrix element $|\langle V \rangle|^2$) with another particle into other momentum states, while the second term describes the reverse gain as two quasiparticles scatter into states $1$ and $2$. The collision integral vanishes if $f$ is equal to the Fermi-Dirac distribution $f_{\rm FD}({\bf p})=1/(\exp(\beta(\varepsilon_{\bf p} - \mu))+1)$ with $\beta=1/k_BT$ the inverse temperature and $\mu$ the chemical potential.
We consider the relaxation of small deviations from this equilibrium distribution, which we parametrize in terms of a deviation function $\psi$ as
\begin{equation}
\delta f({\bf p}) = f({\bf p}) - f_{\rm FD}({\bf p}) = f_{\rm FD} (1-f_{\rm FD}) \psi({\bf p}) \label{eq:deviation} .
\end{equation}
Eigenmodes are then eigenvectors of the linearized collision integral, the eigenvalues of which set their decay rate. Formally,
\begin{equation}
\hat{\cal L}[\psi({\bf p})] = \frac{- \hat{\cal J}[f({\bf p})]}{f_{\rm FD}(p) \bigl(1-f_{\rm FD}(p)\bigr)}
= \gamma \psi({\bf p}) , \label{eq:eigenvalue}
\end{equation}
and, in explicit form,
\begin{align}
&\hat{\cal J}[\psi({\bf p}_1)] = - \frac{m^* V_\star^2}{4\pi \hbar^3} \int \frac{d{\bf p}_2}{(2\pi\hbar)^2} \int d\Omega \, \nonumber\\
&\times f_{\rm FD}({\bf p}_1) f_{\rm FD}({\bf p}_2) \bigl(1 - f_{\rm FD}({\bf p}_{1'})\bigr) \bigl(1 - f_{\rm FD}({\bf p}_{2'})\bigr) \nonumber \\
& \times\Bigl[ \psi({\bf p}_1) + \psi({\bf p}_2) - \psi({\bf p}_{1'}) - \psi({\bf p}_{2'})\Bigr] ,
\label{eq:linearizedcollision}
\end{align}
where $\Omega$ is the angle between the ingoing and outgoing relative scattering momenta. For generality, we assume a constant scattering matrix element $\langle V \rangle = V_\star$, although it is straightforward to extended our calculation to more complex cases. 

For the parabolic dispersion, which has rotational invariance, eigenvectors are labeled by the angular mode $m$ using the decomposition $\psi_m(p) = \int_0^{2\pi} \frac{d\theta}{2\pi} e^{-im\theta} \psi({\bf p})$ with $\theta$ the in-plane angle of ${\bf p}$, where the $\psi_m$ carry a residual dependence on the deviation from the Fermi momentum. The even and odd $m$ describe parity-even and parity-odd Fermi surface deviations sketched in Fig.~\ref{fig:1}(c), and each angular momentum channel has its own decay rate~$\gamma_m$. Note that the effect will remain for a general parity-even dispersion, such as graphene or lattice problems.

Even in linearized form~\eqref{eq:linearizedcollision}, solving the collision integral is a daunting task, hence it is often replaced by a simple relaxation-time ansatz~\cite{baym04,pines66,giuliani05}. Results for transport coefficients were obtained by~\textcite{abrikosov57} in the asymptotic limit where all momenta are precisely fixed at the Fermi surface, which reduces the linearized collision integral to a one-dimensional integral equation for specific perturbations. This equation was solved in~\cite{brooker68,jensen68}, but since the method crucially relies on the zero-temperature limit, it is not apparent how to generalize it to finite temperatures in a systematic way. It is in particular not suitable to describe subleading-in-temperature effects such as the odd-parity decay as well as the onset temperature of such scaling. 
In lieu of a systematic evaluation of the collision integral in the presence of Pauli blocking, there is currently significant effort to develop novel numerical methods, for example, by computing the dynamics of Fermi surface patches~\cite{kryhin21}.
An alternative for transport calculations are diagrammatic approaches, although parity-dependent effects that motivate the current study are so far not possible to analyze with standard Green's function methods, which do not resolve the angular dependence of perturbations.

The technical innovation of our work is an efficient solution of the linearized collision integral~\eqref{eq:linearizedcollision} using an eigenfunction expansion 
in terms of polynomials that are orthogonal with respect to an inner product induced by the Fermi-Dirac distribution
\begin{eqnarray}
\langle g , h \rangle &=& \lambda_T^2 \int \frac{d^2p}{(2\pi\hbar)^2} \, f_{\rm FD}({\bf p}) \bigl(1-f_{\rm FD}({\bf p})\bigr) \, g^*({\bf p}) h({\bf p}) \nonumber \\
&=& \int_{- \beta\mu}^\infty \frac{dw}{4 \cosh^2(w/2)} g^*(w) h(w) , \label{eq:innerproduct}
\end{eqnarray}
where the second line holds for isotropic functions~$g$ and~$h$, and we define the thermal wave length \mbox{$\lambda_T = \sqrt{2 \pi \hbar^2 \beta/m^*}$}, as well as a dimensionless excitation energy \mbox{$w = \beta(p^2/2m^*-\mu)$}. Note that the the linearized collision integral is a Hermitian and positive semidefinite operator with respect to the inner product~\eqref{eq:innerproduct}, with real and nonnegative eigenvalues $\gamma$ (a proof proceeds as for the classical collision integral, see~\cite{mclennan89}). We determine the quasiparticle lifetime by expanding
\begin{equation}
\psi_m(p) = \sum_{\ell=1}^N c_\ell u_\ell(w)
\end{equation}
up to a certain base dimension $N$, where the basis polynomials are chosen orthonormal with respect to~\eqref{eq:innerproduct}, $\langle u_i, u_j \rangle = \delta_{ij}$, with an initial choice $u_1(w) = 1$. We then numerically evaluate the matrix of scalar products  (with details given below)
\begin{equation}
{\cal M}^{\alpha\beta}_{m} = \langle u_\alpha e^{im\theta} , \hat{\cal L} u_\beta e^{im\theta}\rangle ,
\label{eq:Mij}
\end{equation}
the lowest eigenvalue $\gamma_m^{(N)}$ of which provides a variational upper bound (due to the truncation to a finite basis) on the decay rate. Indeed, as shown below, we find rapid convergence at all temperatures with only a small number of basis polynomials. In the following, we systematically compute the decay rates for the first twenty angular Fermi surface deformations and quantify the parity-dependence discussed in the introduction, although the method itself has broader applications.

\begin{figure}[t!]
\scalebox{1.4}{\includegraphics{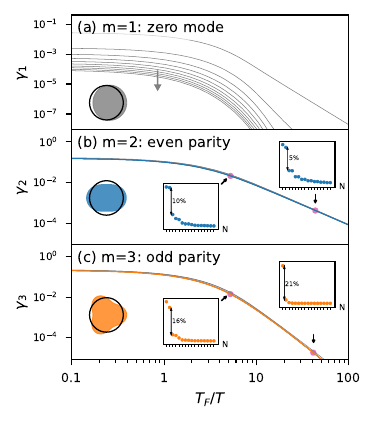}}
\caption{Decay rate (in units of $m^{*2} V_\star^2 T_F/\hbar^5$) for the (a)~\mbox{$m=1$}, (b) \mbox{$m=2$}, and (c) \mbox{$m=3$} modes as a function of inverse temperature obtained by computing eigenvalues of the linearized collision integral~\eqref{eq:linearizedcollision} with an increasing number $N$ of basis functions. While the $m=1$ zero mode converges to zero with increasing $N$, higher modes converge rapidly to a finite value (as illustrated in the insets for selected points on a linear scale as a function of $N$). There is a clear suppression of the low-temperature decay rate for the odd-parity mode $m=3$ compared to the even-parity $m=2$ mode.}
\label{fig:2}
\end{figure}

Consider first the low-temperature basis polynomials, where \mbox{$\beta\mu \to \infty$}. Here, we obtain the orthogonal polynomials in closed analytical form:
\begin{eqnarray}
u_1(w) &=& 1 , \\
u_2(w) &=& \frac{\sqrt{3}}{\pi } w , \\
u_3(w) &=& \frac{3 \sqrt{5}}{4 \pi ^2} w^2-\frac{\sqrt{5}}{4} , \\ 
u_4(w) &=& \frac{5 \sqrt{7}}{12 \pi ^3} w^3-\frac{7 \sqrt{7}}{12 \pi } w ,
\end{eqnarray}
and so on. The first basis function $u_1$ describes a rigid deformation of the Fermi surface and can be thought of as a shift in the chemical potential of the Fermi-Dirac distribution [cf. Fig.~\ref{fig:1}(c)], while higher-order terms account for a smoothing of the Fermi edge. The parametrization in terms of $w$ ensures that higher-order polynomials give a progressively smaller contribution to the eigenvalue, thus illustrating the rapid convergence at low temperatures. An analytical expression for the finite-temperature basis polynomials exists, which mixes even and odd powers of~$u$, but it is also straightforward to generate them via a numerical Gram-Schmidt procedure. A special situation arises at high temperatures \mbox{$\beta\mu \to -\infty$}, where the Pauli blocking factor is irrelevant and the weight function in Eq.~\eqref{eq:innerproduct} reduces to a Maxwell-Boltzmann exponential form. The orthogonal polynomials are then Laguerre polynomials  (in this context called Sonine polynomials~\cite{chapman95}),
\begin{equation}
u_\ell(w) = e^{-\beta\mu/2} L_\ell(w+\beta \mu) . \label{eq:laguerre}
\end{equation}
This reproduces the standard description of a classical two-dimensional gas, for which basis expansion techniques based on a Laguerre expansion have been developed, among others, by 
Enskog~\cite{enskog17}, Burnett~\cite{burnett35}, Grad~\cite{grad49}, and reviewed by Chapman and Cowling~\cite{chapman95}. The framework presented in this Letter is an extension of these high-temperature techniques to all temperatures by including the full extent of Pauli blocking at low temperatures.

In evaluating the matrix elements ${\cal M}_m^{\alpha\beta}$ in Eq.~\eqref{eq:Mij}, we account for energy and momentum conservation by transforming to a center-of-mass ${\bf P}$ and relative momentum ${\bf q}$ for the ingoing scattering states, i.e., ${\bf p}_{1/2} = {\bf P}/2 \pm {\bf q}$ and ${\bf p}_{1'/2'} = {\bf P}/2 \pm {\bf q}'$, which leaves an integration over the angle $\Omega$ between the initial and scattered relative momenta ${\bf q}$ and ${\bf q}'$.  
The resulting four-dimensional integral is computed using a multidimensional adaptive Monte Carlo method~\cite{hahn05}, where we rescale all momenta and energies in units of the thermal wave length $\lambda_T$ to avoid sharp peaks in the integrand.
The chemical potential and temperature are linked by $\beta \mu(T) = \ln (e^{T_F/T} - 1)$.

To establish convergence of our calculations, we show in Fig.~\ref{fig:2} results for the decay rate $\gamma_m^{(N)}$ as a function of $T_F/T$ for three different angular harmonics (a) \mbox{$m=1$}, (b) \mbox{$m=2$}, and (c) \mbox{$m=3$}. Shown are several curves obtained with an increasing number of basis functions $N$. 
First, we show in Fig.~\ref{fig:2}(a) the decay rate $\gamma_{m=1}$, where different lines from top to bottom indicate results obtained with an increasing basis dimension $N=1,\ldots,10$. The $m=1$ mode is special since it is a zero mode of the collision integral, i.e., a conserved mode that is not degraded by electron scattering. For electron-electron interactions, these zero modes are associated with charge (\mbox{$m=0$}) and current (\mbox{$m=\pm1$}) conservation, and correspond to a compression of the Fermi surface and a rigid shift, respectively [cf. Fig.~\ref{fig:1}(c)]. While it is possible to check the cancellation in the collision integral for the \mbox{$m=0$} mode explicitly (where it follows from particle number and energy conservation, which implies \mbox{${\cal M}_{m=0}^{\ell 1} = {\cal M}_{m=0}^{\ell 2} = 0$} for all \mbox{$\ell=1,\ldots,N$}), this is far from obvious for the $m=1$ mode, for which a cancellation between different orders has to occur. Indeed, this is the behavior exhibited in Fig.~\ref{fig:2}(a), which does not converge to a finite nonzero result but rapidly drops to zero as the number of basis functions is increased (indicated by the gray arrow in the figure), with an especially rapid drop at low temperatures. 
Establishing here a cancellation to all orders and recovering the exact zero mode is a strong check of our numerical procedure.

Next, in Figs.~\ref{fig:2}(b) and~(c) we show the first modes not associated with zero modes of the collision integral, namely $m=2$ (parity even) and $m=3$ (parity odd). The corresponding Fermi surface deformation is sketched in the inset of the figures. 
 Gray lines indicate results obtained for basis dimension $N=1,\ldots,15$, and we mark the largest dimension $N=16$ as a blue and orange line, respectively. 
 As is apparent from the figures, these modes behave very differently from the zero modes, and converge rapidly to a finite decay rate at all temperatures, even for a small number of (two or three) basis functions---in fact, the colored lines almost completely cover the gray lines for most temperatures. 
 To illustrate the convergence further, we show in the insets of the figure results on a linear scale at isolated temperatures
 as a function of the basis dimension $N$, where we indicate the size of the largest relative jump explicitly. 
 Note that even though convergence is fast, an accurate result requires going beyond the leading-order rigid Fermi surface deformation at all temperatures (i.e., include basis functions beyond \mbox{$N=1$}). These results establish that the basis expansion is robust and accurate at all temperatures, which is one of the main results of this Letter. 
In addition, comparing Figs.~\ref{fig:2}(b) and (c), it is immediately apparent that while at high temperatures $T \gtrsim T_F$ both decay rates are of comparable magnitude, the odd-parity \mbox{$m=3$} mode drops much more quickly at low temperatures compared to the even-parity \mbox{$m=2$} mode.

\begin{figure}[t!]
\scalebox{0.9}{\includegraphics{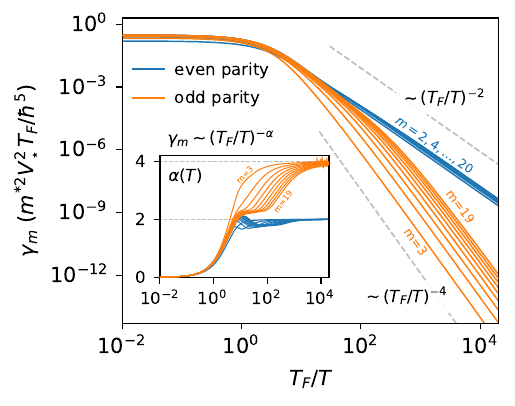}\quad}
\caption{Decay rate $\gamma_m$ of Fermi surface deformations for the first twenty angular harmonics $m=2,\ldots,20$, where blue lines indicate even-parity modes and orange lines odd-parity modes. While even-parity modes display a standard Fermi liquid scaling $T^{-2}$ at low temperatures, odd-parity modes have a significantly reduced decay rate with a $T^{-4}$ scaling at very low temperatures. Inset: Logarithmic derivative of the decay rate $\gamma_m$ to extract the local scaling exponent as a function of temperature. For odd $m$, the crossover from a $T^{-2}$ to a $T^{-4}$ scaling is clearly visible, where the crossover temperature $T_\star$ decreases with increasing~$m$.}
\label{fig:3}
\end{figure}

Let us now analyze this low-temperature scaling in more detail. Figure~\ref{fig:3} shows a double-logarithmic plot of the decay rates over a large temperature range for the first twenty angular harmonics $m$ as a function of~$T_F/T$. The calculations are performed using the first three basis functions, where even-parity modes are shown as blue lines and odd-parity modes as orange lines. The inset of the figure shows the logarithmic derivative of the results in the main plot with respect to $T_F/T$ to extract a local scaling exponent \mbox{$\gamma_m \sim (T_F/T)^{-\alpha(T)}$}. At high temperatures \mbox{$T\gtrsim T_F$}, the decay rate is independent of the angular parameter and approximately constant. This changes at smaller temperatures \mbox{$T\lesssim T_F$}, where Pauli blocking effects start to enter, which reduce the decay rate. For even-parity perturbations, the decay rates cross over to the canonical Fermi liquid \mbox{$\gamma_{m, {\rm even}} \sim (T_F/T)^{-2}$} scaling, which persists over the complete low-temperature range for $T\lesssim0.2 T_F$. Odd-parity modes (orange lines), by contrast, are seen to decay much more slowly, with a rate \mbox{$\gamma_{m, {\rm odd}} \sim (T_F/T)^{-4}$} at the very lowest temperatures. This result agrees with the analytical prediction~\cite{ledwith19}. Different from the even-parity modes, the onset temperature below which the asymptotic low-temperature scaling form is reached depends on the angular parameter $m$, where larger modes first assume a plateau near \mbox{$\alpha\approx 2$} before they cross over to the more rapid decay with \mbox{$\alpha\approx 4$}.
Using \mbox{$\alpha(T_\star)=3$} to mark the onset of the low-temperature scaling, we find \mbox{$T_\star(m=3) = 0.12 \, T_F$}, \mbox{$T_\star(m=5) = 0.04 \, T_F$}, and \mbox{$T_\star(m=7) = 0.02 \, T_F$}. A fit to higher (odd) angular momenta up to $m=19$ gives \mbox{$T_\star(m)/T_F = 0.91/m^2$}. Our calculations using a parabolic dispersion apply directly to doped semiconductors and should be a good approximation for graphene or bilayer graphene at temperatures $T\lesssim T_F$, where the valence band is not depleted by thermal excitations. In particular, for GaAs, a wide hydrodynamic regime is predicted in the range \mbox{$T=1-40$K} despite seemingly weak electron interactions~\cite{ahn22}. With Fermi temperatures in the range \mbox{$T_F=50-200$K} for typical doping densities, we thus predict the anomalous odd-parity transport in a large window of temperatures \mbox{$T \lesssim 5-20$K}, well within the range of current experiments~\cite{braem18}.

In summary, we obtain a complete description of decay rates in two-dimensional Fermi liquids valid at all temperatures by developing a numerically exact solution of the quasiparticle collision integral. 
In particular, we provide an exhaustive characterization of a transport regime in which odd-parity perturbations of the Fermi surface decay anomalously slowly, where the crossover to this new regime occurs at readily accessible temperatures $T\lesssim0.1 T_F$. 
Going forward, the formalism developed here will allow a precise description of 2D Fermi liquids in terms of transport coefficients and collective modes. The 2DEG discussed here is an important proof-of-principle  evaluation of the finite-temperature collision integral, and points the way to broader applications, for example, to graphene or general lattice problems~\cite{deng13}.

\begin{acknowledgments}
We thank M.~Granath, S.~Lara-Avila, H.~Linander, E.~Nilsson, and H.~Rostami for discussions. This work is supported by Vetenskapsr\aa det (Grant No. 2020-04239).
\end{acknowledgments}

\bibliography{bib_collision}

\end{document}